\documentclass[aps,PRResearch,twocolumn,superscriptaddress]{revtex4}
\usepackage{graphicx}
\usepackage{amsmath,amssymb}
\usepackage{bm}
\usepackage{hyperref}

\begin{document}
\title{Robust localized zero-energy modes from locally embedded PT-symmetric defects}
\author{Fatemeh Mostafavi}
\affiliation{Department of Physics and Astronomy, University of Texas Rio Grande Valley, Edinburg, TX 78539, USA}

\author{Cem Yuce}
\affiliation{Department of Physics, Eskisehir Technical University, Eskisehir, Turkey}

\author{Omar S. Magan\~{a}-Loaiza}
\affiliation{Department of Physics and Astronomy, Louisiana State University, 202 Nicholson Hall, Baton Rouge, LA 70803, USA}

\author{Henning Schomerus}
\affiliation{Department of Physics, Lancaster University, Lancaster LA1 4YB,
  United Kingdom}
\author{ Hamidreza Ramezani}
\email{hamidreza.ramezani@utrgv.edu}
\affiliation{Department of Physics and Astronomy, University of Texas Rio Grande Valley, Edinburg, TX 78539, USA}

	\begin{abstract}
We demonstrate the creation of robust localized zero-energy states that are induced into topologically trivial systems by insertion of a PT-symmetric defect with local gain and loss.
A pair of robust localized states induced by the defect turns into zero-energy modes when the gain-loss contrast exceeds a threshold, at which the defect states encounter an exceptional point.
 Our approach can be used to obtain robust lasing or perfectly absorbing modes in any part of the system.
\end{abstract}

\maketitle

\emph{Introduction}.---Interest in topological bandstructure systems \cite{Asb15} has widely expanded into different areas of physics, chemistry, and engineering, including condensed matter physics \cite{Has10,Qi11}, photonics \cite{Lu14,Oza19}, Floquet systems and quantum walks \cite{Kit12,Rec13b}, ultracold atomic gases \cite{Gol16}, acoustics \cite{Yan15}, mechanics and robotics \cite{Kan14,Hub16,Gha19}, electronics \cite{Lee18}, and chemical thin films \cite{Yue14}.

This widespread study of topological systems originated from the classification of the Hermitian topological system \cite{Kit09}. Currently, nontrivial extensions of closed topological systems to their open counterparts attract increasing interest, connecting this area to non-Hermitian concepts such as parity-time (PT) symmetry \cite{ElG18}, and resulting in novel topological applications and phenomena such as topological mode selection and lasing \cite{Sch13,Pol15,StJ17,Zha18,Par18,Ota18}.

Whilst not completely settled, an understanding of such genuinely non-Hermitian topological effects is also emerging \cite{Kun18,Car18,Xio18,Yao18,Yin18,Lee19,Ber19,Imu19,Son19,Lon19,Oku20}, where one has to account for a much larger range of possible symmetries and resulting universality classes \cite{Lie18,Gon18,Kaw19}.
A major complication in these endeavors is the break-down
of the conventional bulk-boundary principle. In particular, a range of studies have identified non-Hermitian degeneracies known as exceptional points  (EPs) as a mechanism to create robust defect states, even when starting from systems that are trivial in their Hermitian limit
\cite{Zhu14,Mal15,Ge17,Mal18}.

These observations suggest that the boundary at the interface of two non-Hermitian systems can be enough to induce a topological transition, even when the coupling configuration in the bulk---which completely determines the topological phase in Hermitian systems---does not change.
Nonetheless, so far, most studies of this phenomenon still utilized systems that either already possessed topological states in the Hermitian limit \cite{Sch13,Pol15,Zeu15,Wei17,Jin17,Lie18,Pan18,Yuc19}, or altered the coupling configuration in some suitable way  \cite{Zhu14,Mal15,Ge17,Mal18,Lan18,Ni18}.

In this work, we demonstrate that a PT-symmetric defect embedded into the topologically trivial phase of a Hermitian system is indeed sufficient to create localized symmetry-protected defect states.
The creation of these states is again manifested by an EP, and the states reside in the band gap, as desired for many applications \cite{Aka03,Son05}. In particular, single site non-Hermitian defects are good candidates for designing conventional photonic crystal lasers \cite{Pai99,Nod01,Col03,Zho16}, and a wide range of other applications such as strain field traps \cite{Sie88} and strong photon localization \cite{Joh87}.

Utilizing a PT-symmetric defect has a range of additional benefits. The PT-symmetry facilitates the emergence of exceptional points, which have been used to control lasing emission \cite{Lie12}, enhance sensing \cite{Wie14,Che17}, create coherent perfect absorption \cite{Cho10}, and can induce directional transport \cite{Ram10} and conical diffraction \cite{Ram12}. However, the relation of these effects to topological transitions have not been addressed in these studies. Given that we demonstrate the appearance of the defect states without any change of the coupling configuration, our study paves the path to create robust localized states on demand and at any part of the lattice. This widens the scope for practical applications in quantum sensing, topological memories, and topological lasing, where one might desire to create or eliminate a robust localized zero mode at any location within a given structure.

 \begin{figure}
 	\centering
 	\includegraphics[width=\columnwidth]{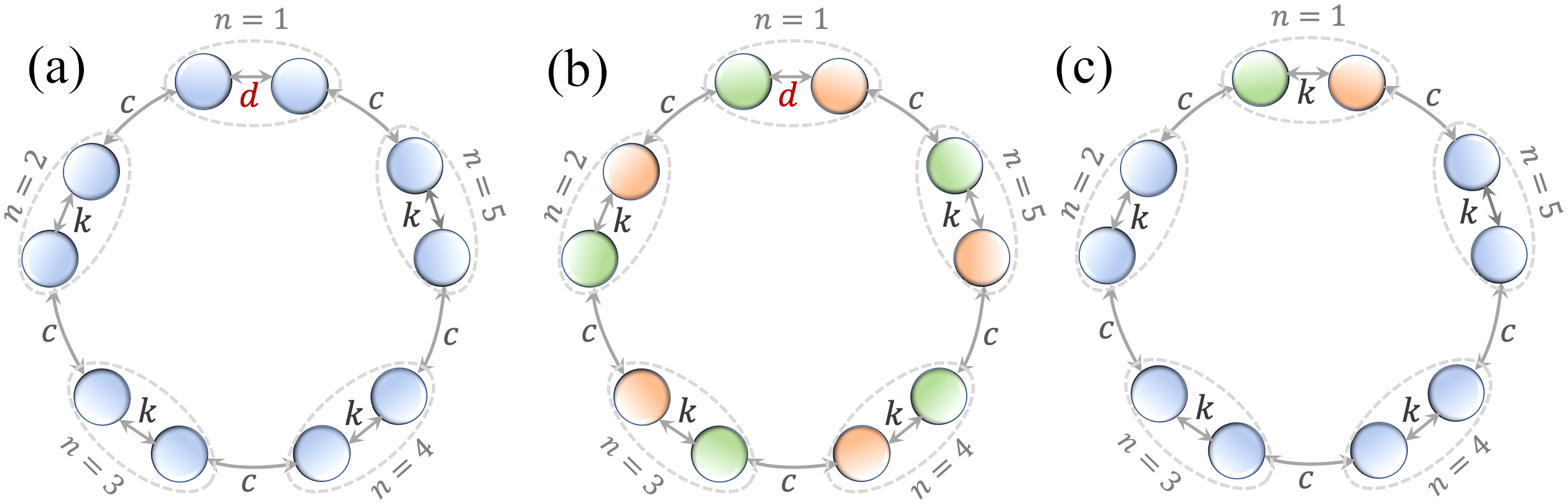}
 	\caption{Sketch of coupled-dimer resonators lattices with an embedded defect. (a) A Hermitian lattice with ${\cal N}=10$ resonators (light blue circles), hence $N=5$ dimers, where all the resonators are passive with no net gain or loss. Resonators in the same dimer are coupled by the intra-dimer coupling strength $k$, while the inter-dimer coupling strength $c$ is assumed to fulfill $c<k$.
 The defect is created by changing the intra-dimer coupling strength in a single dimer, denoted as $n=1$, from $k$ to $d$. In this system, the defect modes are always hybridized, and never become zero modes.
(b) PT-symmetric version of the lattice of panel (a), where the orange (green) circles depict resonators with the gain (loss) of strength $\gamma$. We show that this lattice can support localized zero modes, which emerge through an exceptional point when $d$ is sufficiently small and $\gamma$ sufficiently large.
(c) Modified set-up in which the PT-symmetric defect is embedded into the Hermitian system. This lattice can  support the same type of zero modes as the model in panel (b), which demonstrates that such modes can be induced by embedding a non-Hermitian defect into a topologically trivial Hermitian system.}
 	\label{fig1}
 \end{figure}

With such practical applications in mind, we demonstrate the creation of these states for a specific structure of experimental interest, namely, a periodic dimer chain that in the passive case corresponds to a Su-Schrieffer-Heeger (SSH) chain \cite{Su79} in its trivial coupling configuration. In its topologically nontrivial counterpart configuration, topological lasing utilizing edges or interfaces has been demonstrated in a number of studies \cite{StJ17,Zha18,Par18,Ota18}. In particular, in Ref.~\cite{StJ17}, the lasing of an SSH edge state was facilitated by pumping the system only at the edge. In contrast, we create a localized defect mode suitable for lasing inside the trivial  phase, by only utilizing the non-Hermiticity induced by gain and loss. This long-living state is pinned to the centre of the band gap, and, as is typical for states emerging in EPs, is accompanied by a second robust mode of shorter life time.

\emph{Model}.---We consider a non-Hermitian one-dimensional dimer lattice with periodic boundary conditions, representing, e.g., evanescently coupled microdisk resonators \cite{Cao15} as shown in Fig.~\ref{fig1}. While non-Hermiticity can be obtained in different ways, we consider the case where the real part of the resonance frequency of the coupled modes is $\omega_0$, while the imaginary part $\gamma$
(with $\gamma>0$  representing gain  and $\gamma<0$ representing loss) in each dimer unit cell is antisymmetric.
The coupled-mode equations that describe the dynamics in this lattice are given by
\begin{equation}
\begin{array}{c}
i\partial_{t} \psi_{n}= -k \varphi_{n} - c \varphi_{n-1} +i\gamma \psi_{n}\\
i \partial_{t} \varphi_{n}=- k \psi_{n} - c \psi_{n+1} -i\gamma \varphi_{n}
\end{array}
\label{eq1}
\end{equation}
where $\psi_{n}$ and $\varphi_n$ are the modal field amplitudes in the $n$th gain and loss disk. We assume that the  intra-dimer couplings $k$ and inter-dimer coupling $c$ between the adjacent disks are real and fulfill $k>c$, and without loss of generality set $\omega_0=0$. Periodic boundary condition are obtained by requiring $\vec{\Psi}_{N+n}\equiv(\psi_{N+n}\quad \varphi_{N+n})^T=\vec{\Psi}_n\equiv(\psi_{n}\quad \varphi_{n})^T$, with $N$ being the total number of dimers in the lattice.
\begin{figure}
	\includegraphics[width=\columnwidth]{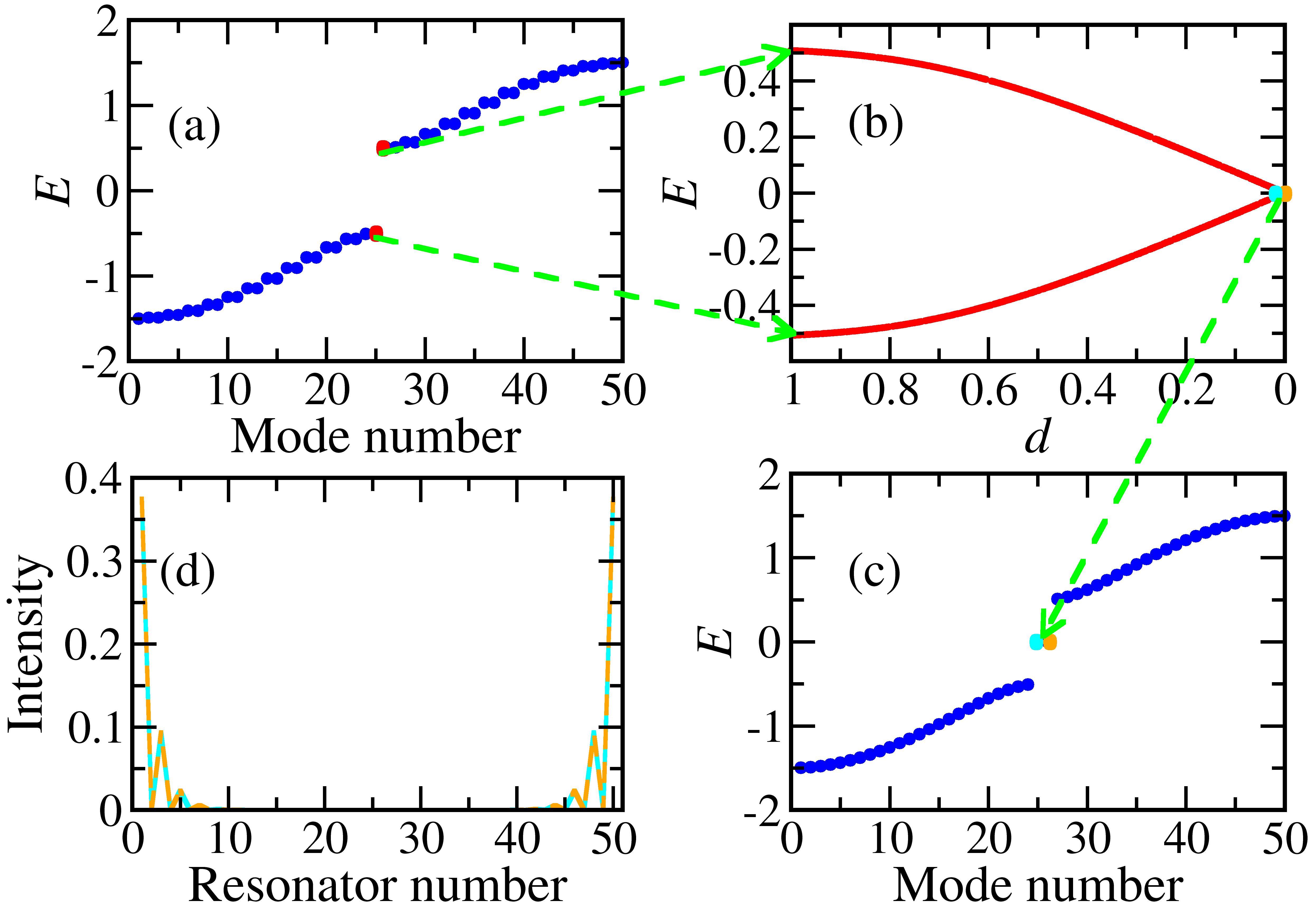}
	\caption{Defect modes in the  Hermitian model of Fig.\ref{fig1}(a). (a) Quantized band structure for a system of 100 resonators with couplings $c=0.5$ and $k=1$, and no defect, $d=k=1$.
(b) Changing the defect coupling to $d<k$ moves two modes from the band edges  (identified by the red dots in (a)) into the gap.
(c) Energy spectrum for $d=0$, upon which the two sites on the defect dimer
become the edges of a system with open boundary conditions.
(d) The two modes always remain weakly hybridized, with mode profiles that are localized symmetrically on both of these effective edges.
Note that in this and the following representations of the mode profiles, resonators are numbered so that the first and last resonators are those of the defect dimer.
}
	\label{fig2}
\end{figure}
\begin{figure}
	\includegraphics[width=\columnwidth]{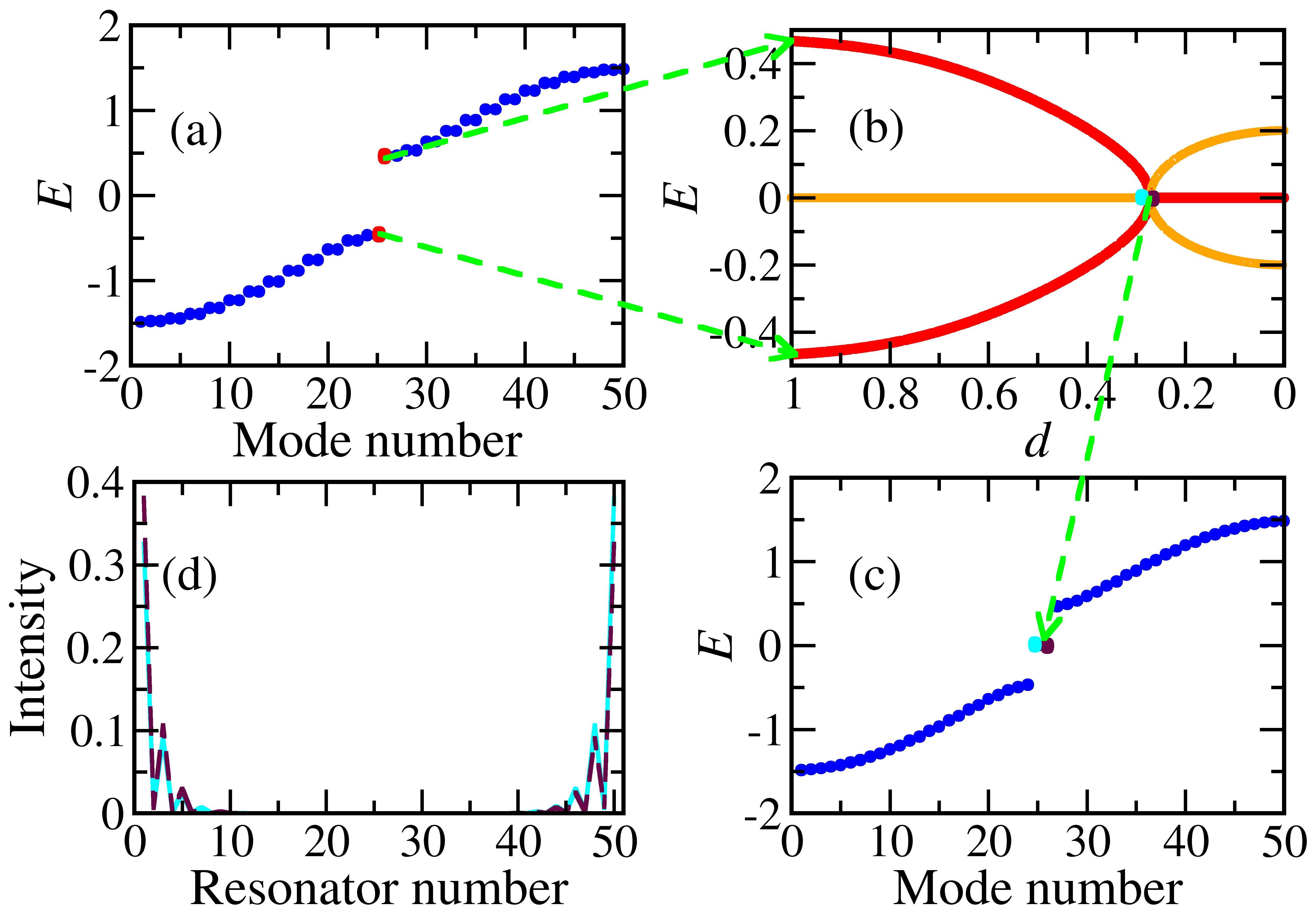}
	\caption{Defect modes in the  non-Hermitian model of Fig.~\ref{fig1}(b), in analogy to Fig.~\ref{fig2} but with a finite gain-loss parameter $\gamma=0.2$.
 (a) Compared to the Hermitian case, the band structure of the system without a defect is similar, but the gap is reduced.
(b) Changing the defect coupling to $d<k$ again creates two defect modes, but these now undergo an additional transition in which they become zero modes with $\mathrm{Re}\, E=0$. This corresponds to an exceptional point, which for the given parameters occurs at $d=0.272$.
 (c) Energy spectrum at the exceptional point.  (d) At the exceptional point the mode profiles are still symmetric, but this symmetry is violated beyond the exceptional point, as shown in Fig.~\ref{fig4}.}
	\label{fig3}
\end{figure}

The Hermitian system corresponds to a periodic variant of the celebrated SSH chain \cite{Su79}. This system possesses a chiral symmetry, which guarantees that the real energy spectrum is symmetric about $E=0$, as well as separate parity and time-reversal symmetries.
In the non-Hermitian case these symmetries are broken, but the  balanced gain and loss makes the system PT symmetric \cite{ElG18}, i.e., the combination of parity and time-reversal still holds.
As a consequence, the complex resonance-energy spectrum is symmetric with respect to
the axis  $\mathrm{Im}\,E=0$, where the occurrence of pairs of complex-conjugated energies signifies the so-called PT-broken phase.
Furthermore, instead of the chiral symmetry the non-Hermitian system displays a non-Hermitian charge-conjugation or particle-hole symmetry \cite{Sch13,Mal15,Ge17}, i.e, the combination of the chiral symmetry with the time-reversal symmetry, which guarantees that the resonance-energy spectrum is  symmetric with respect to the axis $\mathrm{Re}\,E=0$. Notably, this permits the existence of unpaired modes with $\mathrm{Re}\,E=0$, which is the key feature that we will exploit in the following.

For an infinitely long homogeneous system, the band structure of the above model is given by $
E^{(\pm)}(q)=\pm\sqrt{c^{2}+k^{2}-\gamma^{2}+2 c k \cos (q)}$ where $q$ is the Bloch wave number \cite{Ram12b}. For the Hermitian case with $\gamma=0$, schematically depicted in Fig.~\ref{fig1}(a), the two bands are separated by a gap of size $\Delta_H=2(k-c)$. This gap closes when $k=c$, when the chain becomes non-dimerized, signalling a band inversion as one passes from one topologically distinct coupling configuration to the another. For nonzero value of $\gamma$ [Fig.~\ref{fig1}(b)] the gap size reduces to $\Delta_{NH}=2\sqrt{(k-c)^2-\gamma^2}$, which for small values of $\gamma$ can be approximated as $\Delta_{NH}\approx \Delta_H-\frac{\gamma^2}{k-c}+O(\gamma^3)$.  Therefore, the gap for the non-Hermitian lattice is smaller than the gap for the Hermitian lattice. In particular, the gap becomes zero at $\gamma=\gamma_{EP}=k-c$, where the first two modes at the edge of the Brillouin zone merge with each other in an exceptional point. For $k-c<\gamma<k+c$, a part of the dispersion is purely imaginary, corresponding to states with degenerate resonance frequency $\mathrm{Re}\,E^{(\pm)}(q)=0$. For $\gamma=k+c$ all the modes from both bands have merged, creating a flat band of states with different life times \cite{Qi18,Ram17}.
The same features hold for the finite system, where the periodic boundary conditions lead to the quantization $q=q_m=\frac{2\pi m}{N}$ of Bloch wave number. This gives rise to discrete resonance frequencies $E_m^{(\pm)}=E^{(\pm)}(q_m)$, where $m=0,1,2,\ldots N$ is the index of the associated super-mode.

At this point let us assume that we can adiabatically change the value of one of the intra-dimer couplings $k$ to some other value $d<k$ [see Fig.~\ref{fig1}(a)].
This defect continues to preserve the symmetries of the system with exception of translation symmetry, and causes the emergence of two defect modes in the gap.
The situation for the Hermitian case is illustrated in Fig.~\ref{fig2}.
Panel (a) shows the band structure of the periodic system with $d=k$, which is symmetric as dictated by the chiral symmetry, and gapped.
As shown in panel (b), by decreasing the value of $d$ from $k$ toward zero two defect modes appear, which depart from the band edges and move into the gap. Both modes are related by the chiral symmetry, and each mode has a finite weight at both edges of the system. For a finite system, these two defect modes are therefore hybridized edge modes separated by a non-zero gap, and thus not topologically robust.

 \begin{figure}
 	\includegraphics[width=\columnwidth]{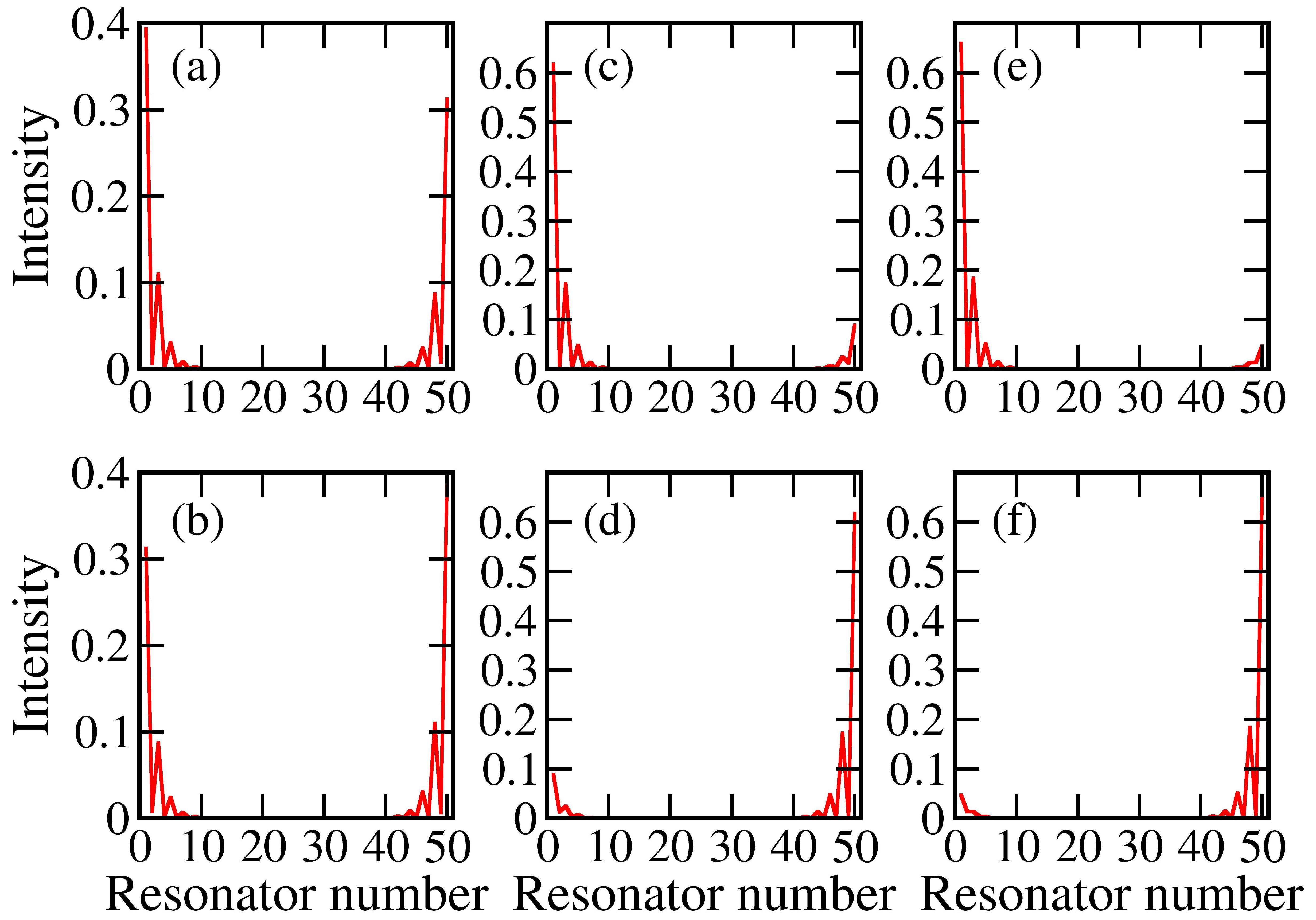}
 	\caption{Mode profiles for the zero modes described in Fig.~\ref{fig3}, but for fixed defect coupling $d=0.27$ with (a,b) $\gamma=0.2$, (c,d) $\gamma=0.3$, and (e,f) $\gamma=0.4$. The upper panels show the zero mode with $\mathrm{Im} \,E>0$, which is preferentially localized on the gain site of the defect dimer, whilst the lower panels show the mode with $\mathrm{Im}\,E<0$, which is preferentially localized on the lossy site. This asymmetry increases for increasing $\gamma$, hence, as one moves deeper into the PT-broken phase.
}
 	\label{fig4}
 \end{figure}

The contrasting situation of the non-Hermitian system with $\gamma\neq 0$ is shown in Fig.~\ref{fig3}. As depicted in panel (a), for $\gamma<k-c$ the band structure of the periodic system ($d=k$) remains symmetric and real, as dictated by PT and particle-hole symmetry, but the gap is smaller than the corresponding Hermitian case. As shown in panel (b), when reducing $d$ below the value of $k$, two defect modes  again emerge from the band edges and move into the gap. However, unlike in the Hermitian case, these modes meet in an exceptional point for a non-zero value of $d$ [Fig.~\ref{fig3}(c)], meaning that the non-Hermitian system can support zero-energy modes without changing the coupling configuration in the bulk of the system.
These zero-energy modes remain exponentially localized around the defect position, and right at the exceptional point are symmetrically localized around the defect, as shown in panel (d).

We note that the position of exceptional points in a PT-symmetric system is not robust, but their existence is---changing parameters, such as introducing disorder, simply shifts the exceptional point to another position in parameter space. Recently robust exceptional points have been proposed in Ref.~\cite{Yuc19b} with application in robust exceptional point sensing, however, those robust exceptional points are not spatially localized and thus may not provide strong feedback for lasing application.
Here, we encounter an exceptional point that signals the emergence of a pair of modes with robust frequency $\mathrm{Re}\,E=0$, in a system in which the bulk band structure remains real and has not undergone a topological transition. At the same time, this represents a mechanism to selectively break the PT symmetry of one predetermined mode in the whole system.

\begin{figure}
	\includegraphics[width=\columnwidth]{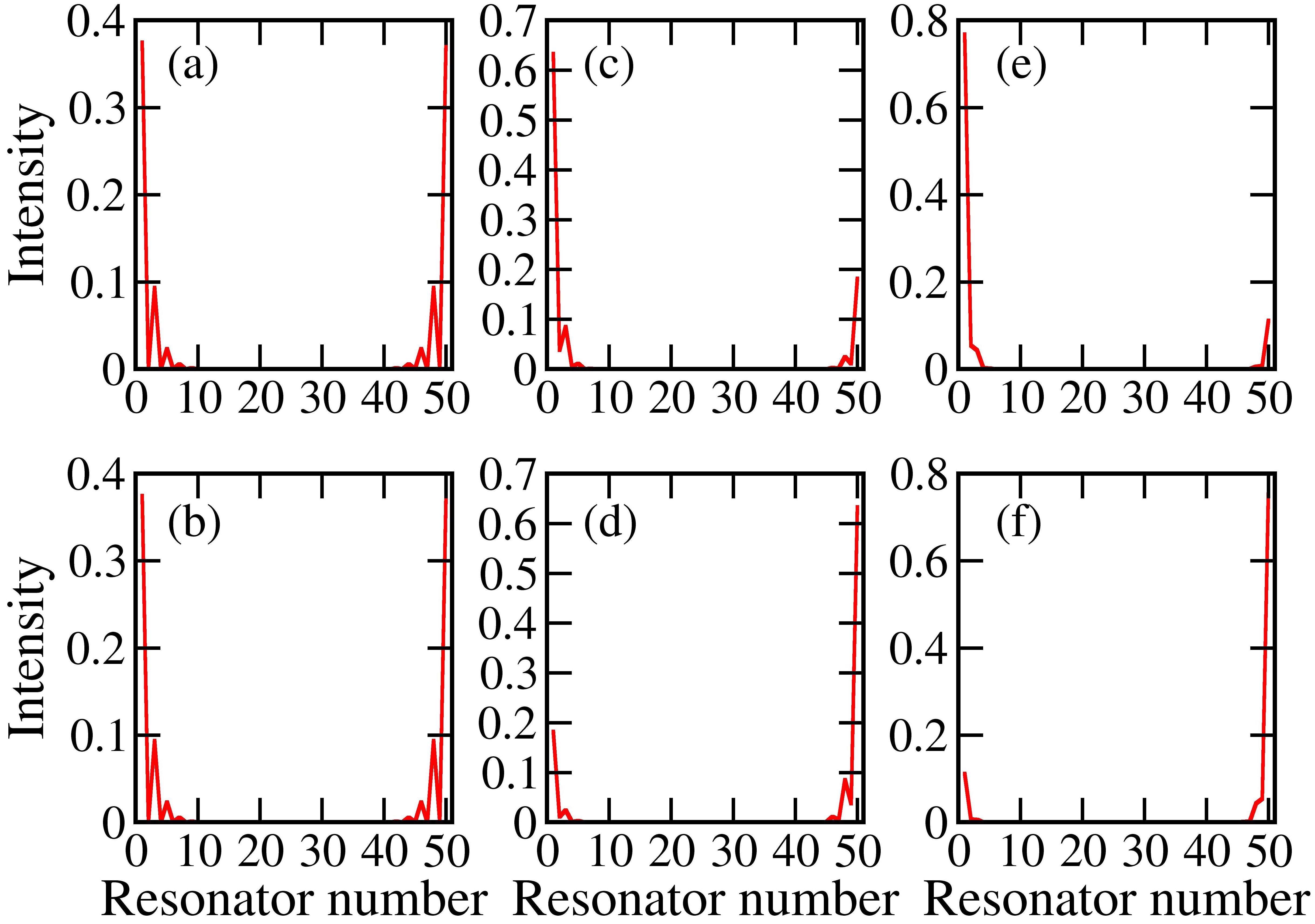}
	\caption{Same as Fig.~\ref{fig4}, but for the model of Fig.~\ref{fig1}(c), where the zero modes are induced by a PT-symmetric defect, without any changes to the coupling configuration (again $c=0.5$ and $k=1$). This transition occurs at an exceptional point at $\gamma=1$; the panels show the modes for
 (a,b) $\gamma=1$, (c,d) $\gamma=1.2$, and (e,f) $\gamma=1.5$.}
	\label{fig5}
\end{figure}

We now continue to characterize these modes in detail as one passes over the exceptional point into the PT-broken phase.
We note that this can be done in two ways, by fixing $\gamma$ and decreasing $d$, or by fixing $d$ and increasing $\gamma$.
In Fig.~\ref{fig4} we show the mode profile for a fixed defect coupling $d=0.27$ and different values $\gamma=0.2,0.3, \text{ and } 0.4$ of the gain and loss parameter, while other parameters of the lattice are the same as  in Fig.~\ref{fig3}. Larger values of $\gamma$ indicate that the system is deeper in the broken phase. The upper panels are associated with the zero-energy modes with a positive imaginary part of their eigenvalues,  $\mathrm{Im}\,E>0$, whilst lower panels show the mode profiles of the mode with a negative imaginary part, $\mathrm{Im}\,E<0$.
As $\gamma$ is increased, these modes become asymmetric, preferentially localized either on the gain site of the defect (for $\mathrm{Im}\,E>0$) or on the corresponding loss site (for $\mathrm{Im}\,E<0$), as is typical for PT-broken states.
We confirmed numerically that these features remain robust under the introduction of disorder in the couplings, up to a threshold that depends on how deep one is situated inside the broken phase. In addition, we found that when keeping $d$ fixed the imaginary part of zero modes changes only weakly with such disorder.

The described mechanism of zero-mode creation translates to a wide class of PT-symmetric defects.
Consider, for instance, a periodic lattice where there is no defect in the couplings, and gain and the loss parameter is zero everywhere except in one unit cell, as schematically depicted in Fig.~\ref{fig1}(c). In practice this means that all the resonators remain passive with no net gain or loss except for the two resonators of one unit cell, one with net gain $\gamma>0$ and the other with net loss $-\gamma$. With this defect embedded into the ring, we find that localized defect states  occur for any finite value of $\gamma$, and that their spectral position moves inside the gap when $\gamma\gtrsim k-c$, mimicking the scenario with gain and loss at the edges of a finite system \cite{Zhu14}. The defect states again turn into zero modes in an exceptional point, which occurs at $\gamma= k$ and hence coincides precisely with the exceptional point for an isolated dimer. Similar to the case with defect coupling, the defect states are spatially symmetric at the exceptional point, but become increasingly asymmetric deeper in broken phase, as shown in Fig.~\ref{fig5}.

\begin{figure}
	\includegraphics[width=\columnwidth]{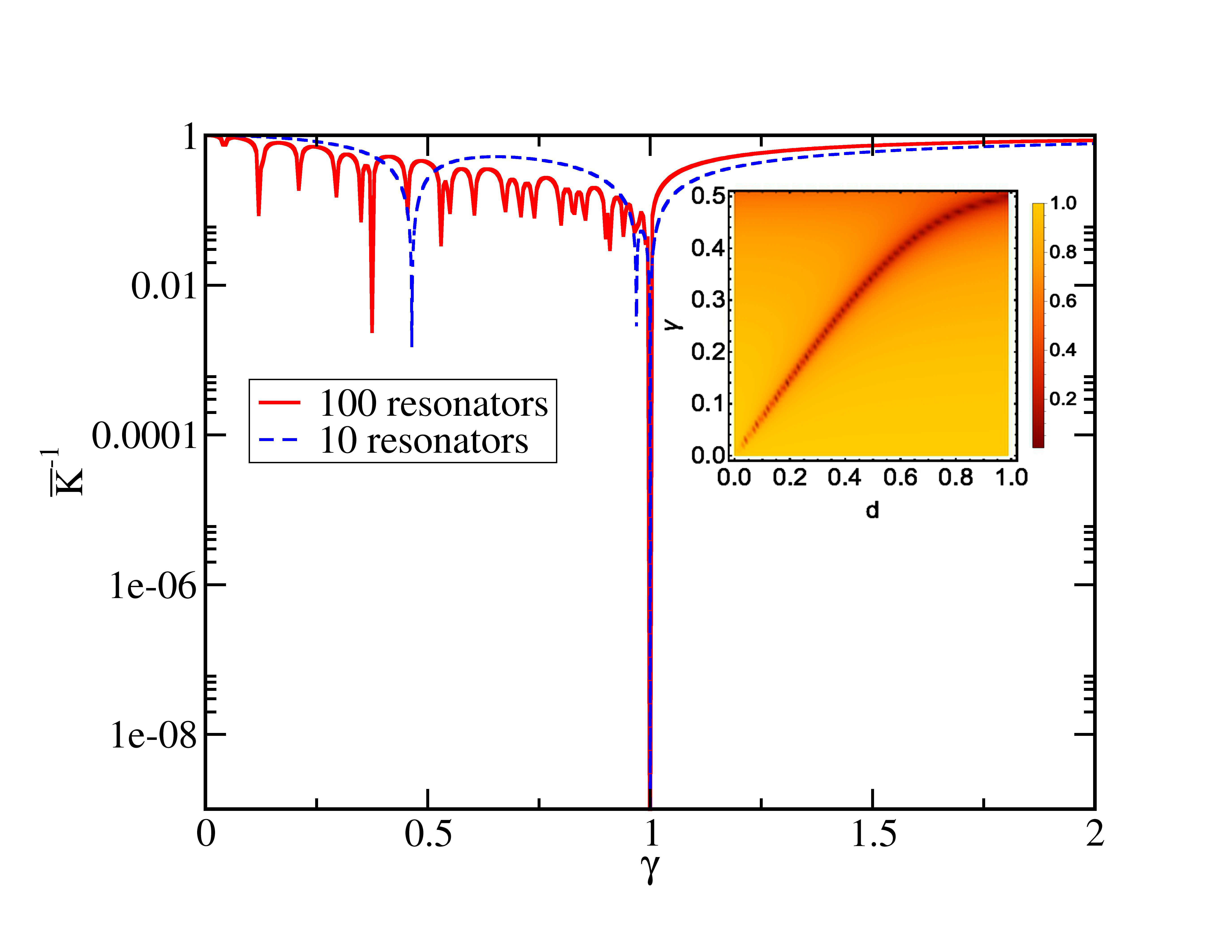}
	\caption{
 Inverse of the mean Petermann factor $\overline{K}$ as an indicator of exceptional points. In the main panel, this indicator is shown as a function of $\gamma$ for the model of Fig.~\ref{fig1}(c), and the two system sizes ${\cal N}=10$ (blue) and ${\cal N}=100$ (red). This confirms the existence of an
exceptional point  at $\gamma=1$, beyond which one encounters the robust zero modes.
For the model of Fig.~\ref{fig1}(b), the inset shows $\overline{K}^{-1}$ as a colour density plot as a function of $\gamma$ and $d$.
As in the other figures, the couplings are fixed to $k=1$ and $c=0.5$. For $d\to 1$, where the defect disappears, the exceptional point mergers with the exceptional point of the bulk band structure, $\gamma_{EP}=k-c=0.5$.\label{fig6}}
\end{figure}

\emph{Eigenfunction analysis}.---To further illuminate the conditions under which the defect states enter the PT-broken phase,
where they become robust zero-energy modes, we turn to the analysis and characterization of the bi-orthogonal set of the eigenvectors \cite{Cha98}.
Let $\langle L_{n}|$ and $|R_{n}\rangle$
denote the left and right eigenvectors corresponding to the eigenvalue ${\cal
E}_n$ of a general non-Hermitian Hamiltonian ${\cal H}$ , i.e.
\begin{equation}
\langle L_n|{\cal H}=\langle L_n|{\cal E}_n, \quad {\cal H}|R_n\rangle= {\cal E}_n|R_n\rangle.
\end{equation}
The vectors can be normalized to satisfy
$\left\langle L_{n}|R_{m}\right\rangle =\delta_{nm}$,
upon which
$\sum_{n}^{\cal N}\left|R_{n} \right\rangle \left\langle L_{n}\right|=1$.
Here ${\cal N}$ is the dimension of the Hilbert space, which for our systems is equivalent to the total number of resonators, ${\cal N}=2N$.

An observable that measures the non-orthogonality of the modes, and can be used to identify the proximity to
the exceptional point in the presence of finite-size effects, is the so-called \emph{Petermann factor}
\begin{equation}
K_{n}=\left\langle L_{n}| L_{n}\right\rangle \left\langle R_{n}|R_{n}\right\rangle,
\end{equation}
which determines the quantum-limited linewidth of lasers \cite{Pet79,Sie89,Pat00}.
At an exceptional point, the eigenvectors associated with the degenerate eigenvalue
coalesce, leading to a Petermann factor that diverges as \cite{Ber03,Lee08}
\begin{equation}
\overline{K}\sim 1/ |\gamma-\gamma_{\cal PT}|
.
\end{equation}

We have studied the averaged Petermann factor
\begin{equation}
\overline{K}=\frac{1}{\cal N} \sum_{n=1}^{\cal N}K_{n}
\end{equation}
which takes the value $1$ if the eigenfunctions of the system are orthogonal, while it is larger than one otherwise.

As shown in the main panel of Fig.~\ref{fig6}, for the model of Fig.~\ref{fig1}(c) the mean Petermann factor indeed diverges at $\gamma=1$, which signals the transition of the defects modes into the symmetry-broken phase.
For the model in Fig.~\ref{fig1}(b), the inset of Fig.~\ref{fig6} shows
how the transition depends on the interplay of $d$ and $\gamma$ in a system of $\mathcal{N}=100$ resonators. In this inset, the red ridge delineates the transition line, so that the  phase with robust localized zero modes is found above this curve.
Therefore, this technique can be used to determine the transition reliably for finite systems as a function of the system parameters.

\emph{Conclusions}.---In summary, we demonstrated that robust zero modes  can appear when a non-Hermitian defect is embedded into the topologically trivial phase of a Hermitian system. Our detailed analysis reveals that these states become robust in an exceptional point that is independent of the bulk structure, and that this phenomenon carries over to a range of PT-symmetric defects. In the form as presented here, the described systems could be realized on a variety of platforms in which SSH models with gain and loss have already been realized \cite{Pol15,Zeu15,Wei17,StJ17,Zha18,Par18,Pan18}.
Our approach can be easily extended to higher dimensions, and could provide useful insights also for the study of disordered systems, for which robust dynamical effects of localized modes have been recently reported \cite{Wei20}.

%

\end{document}